\documentclass[12pt]{article}
\usepackage{a4,amsmath,epsfig,amssymb}
\usepackage[latin1]{inputenc}
\begin{document}

\begin{titlepage}
\begin{flushright}
PCCF RI XX \\
INFN YY \\
\end{flushright}
\renewcommand{\thefootnote}{\fnsymbol{footnote}}
\vspace{1.0em}

\begin{center}
{\bf \Large{Asymmetries in Three-Body Decays of the $B$-Meson}}
\end{center}
\vspace{1.0em}
\begin{center}
\begin{large}
E. Di Salvo$^{1,2}$\footnote{Elvio.Disalvo@ge.infn.it}
 \\
\end{large}
\vspace{2.3em}
%
%
$^{1}$ Laboratoire de Physique Corpusculaire de Clermont-Ferrand \\
IN2P3/CNRS Universit\'e Blaise Pascal \\
F-63177 Aubière Cedex. FRANCE \\ 
$^{2}$ INFN - Sez. di Genova. ITALY
\end{center}
\vspace{5.0em}
\begin{abstract}
\vspace{1.0em}
\noindent
We suggest to analyze the three-body decays $B^{\pm} \to K^{\pm} p {\bar p}$ 
and $B^{\pm} \to K^{\pm} \pi \pi$ by means of two different asymmetries. The 
former asymmetry, a T-odd one, vanishes and is proposed as a test for excluding 
biases in the experimental data. The latter asymmetry, which is T-even, may be 
useful in singling out time reversal odd contributions to the decay, and, 
possibly, hints at physics beyond the standard model.

\vskip 0.5cm
\end{abstract}
%
%

%
%
%
\end{titlepage}
\newpage
\noindent

The physics of the $B$-meson has opened a new door in the sector of 
hadronic weak interactions. In particular, although numerous $B$ decays, 
realized by more than a decade, have confirmed the CKM mechanism for 
CP violation\cite{bs}, it is generally believed that new physics beyond 
the standard model might be hidden in some particular decays of this kind. 
Therefore several experiments in this sense have been suggested or even 
realized. We consider among them the following two pairs of CP-conjugated 
three-body decays of $B$, which offer the possibility of new kinds of tests:  
\begin{equation}
B^{\pm} \to K^{\pm} ~~ p ~~ {\bar p}\cite{ba1,be1} ~~~~ \mbox{and} 
~~~~ B^{\pm} \to K^{\pm} ~~ \pi     ~~ \pi\cite{be2,ba2,ba3}. 
\label{dcys} 
\end{equation}
The former decays present an anomalously large charge 
asymmetry\cite{ba1,be1,pdg}. Moreover the decay $B^+ \to 
K^+ ~ p ~ {\bar p}$ was investigated in the past years by 
means of a Dalitz plot analysis\cite{ba1}, according to the 
suggestion by Rosner\cite{rs}. We propose an alternative type of 
analysis, especially for the new data collected by the LHCb 
detector\cite{cp1,cp2}; this is based on asymmetries related to 
T-odd\cite{va,wf,bl,bdl,bdl1,gr,aj,ajd,ajd1} and 
T-even\cite{at,aj,daj,ajd1} products of momenta or/and angular 
momenta. As we shall see, our method allows also to investigate
the CP asymmetry from a different point of view. 

We schematize the decays (\ref{dcys}) as 
\begin{equation}
B \to a ~~ {\cal S}, ~~~~ \mbox{where} ~~~~ {\cal S} \equiv (b, ~~ K).
\label{gr}
\end{equation}
Moreover $a$ is $p$ (${\bar p}$) and $b$ is ${\bar p}$ ($p$) in the 
former $B^{\pm}$ decay; on the contrary, $a$ and $b$ are the two 
pions in the latter decay. We propose to apply to such decays 
two different tests, based on the determination of asymmetries of 
the type
\begin{equation}
{\cal A} = \frac{N(C>0) - N(C<0)}{N(C>0) + N(C<0)}. 
\label{asm}
\end{equation}
Here $C$ is a correlation made up with independent momenta of the particles involved in the decay; 
moreover, $N(C>0)$ ($N(C<0)$) is the number of events for which the correlation is positive (negative). We consider two different types of 
correlation, a T-odd and a T-even one:
\begin{equation}
C_o = {\bf p}_B \cdot {\bf p}_K \times {\bf p}_a, \ ~~~~~~ \
\ ~~~~~~ \ C_e = {\bf p}_a \cdot {\bf p}_K. \label{cr3}
\end{equation}
Here ${\bf p}_B$ is the momentum of the $B$-meson in the laboratory frame, ${\bf p}_a$ the momentum of $a$ in the rest frame of $B$ ($B$-frame from now on) and ${\bf p}_K$ the momentum of $K$ in the rest frame of ${\cal S}$ (to be named ${\cal S}$-frame). We call ${\cal A}_o$ and ${\cal A}_e$ the two asymmetries associated respectively to the correlations $C_o$ and $C_e$.

Now we study the main features of the asymmetries associated to the two correlations above. As regards the T-odd asymmetry ${\cal A}_o$, 
we observe that $B$ is spinless, therefore, in the $B$-frame, $a$ and ${\cal S}$ are in a helicity eigenstate, associated to a given eigenvalue 
$\lambda_a$ = $\lambda_{\cal S}$ = $\lambda$, where $|\lambda|$ $\leq$ $s_a$ and $s_a$ is the spin of $a$. This implies, in turn, that a rotation around ${\bf p}_a$ in this frame changes the decay amplitude by an overall phase, and therefore the angular distribution of $b$ and $K$ is isotropic around that direction. Obviously we refer to a rotation involving just the variables of the decay process, which is neatly separated from the production process; for example, the rotation does not affect ${\bf p}_B$. In particular, a rotation of this kind, by an angle $\pi$, leaves the distribution unchanged, 
while inverting the sign of the correlation $C_o$. Then the T-odd asymmetry ${\cal A}_o$ is zero for both decays (\ref{dcys}). This result holds true for any three-body decay of a spinless resonance. The asymmetry ${\cal A}_o$ may be usefully employed as a check for singling out possible systematic errors in data samples.

In order to analyze the T-even asymmetry ${\cal A}_e$, we write the decay amplitude in the helicity representation:
\begin{equation}
H_{\lambda_a,\lambda_b} (\theta, \phi) = \sum_j {\cal A}^j_{\lambda_a\lambda_b} d^j_{\lambda_a\lambda_b} (\theta) e^{i\lambda_a\phi}. \label{3mm}
\end{equation}
Here $\lambda_b$ is the helicity of $b$ in the ${\cal S}$-frame, 
$|\lambda_b| \leq s_b$, $s_b$ being the spin of $b$; the dummy index $j$ runs over non-negative integers or half-integers, according to the 
values of $s_a$ and $s_b$. Moreover the $d^j$'s are the usual rotation matrices, $\theta$ being the polar angle between ${\bf p}_K$ and ${\bf p}_a$, while $\phi$ is the azimuthal angle between two planes, singled out by the pairs of vectors (${\bf p}_a$, ${\bf p}_K$) and (${\bf p}_a$, $\hat{\bf n}$) respectively, where $\hat{\bf n}$ is a unit vector normal to the decay plane. Lastly, the ${\cal A}^j_{\lambda_a\lambda_b}$'s read as
\begin{equation}
{\cal A}^j_{\lambda_a\lambda_b} = \frac{\sqrt{2j+1}}{4\pi} a^0_{\lambda_a} c_j b^j_{\lambda_b}. \label{ampam}
\end{equation}
Here the $c_j$'s are such that 
\begin{equation}
|{\cal S}\rangle = \sum_j c_j |{\cal S}_j\rangle,
\end{equation}
where $|{\cal S}_j\rangle$ are angular momentum eigenstates of the $(b, ~ K)$-system, with helicity $\lambda_a$, and $\sum |c_j|^2$ = 1. Moreover $a^0_{\lambda_a}$ and $b^j_{\lambda_b}$ are rotationally invariant amplitudes describing, respectively, the decay $B \to a ~~ {\cal S}$ in the $B$-frame and the process ${\cal S}_j \to b ~~ K$ in the ${\cal S}$-frame. Such amplitudes depend on the rest masses of $B$ and ${\cal S}$
respectively.  

Therefore the decay probability reads as
\begin{equation}
{\cal P} (\theta) =  \sum_{\lambda_a,\lambda_b}\sum_{j,j'} 
{\cal A}^j_{\lambda_a\lambda_b}{\cal A}^{j'*}_{\lambda_a\lambda_b}
d^{j}_{\lambda_a\lambda_b} (\theta) d^{j'}_{\lambda_a\lambda_b} (\theta). \label{dpr}
\end{equation}
Note that this probability is independent of the azimuthal angle
$\phi$, coherently with the vanishing of the T-odd asymmetry. 
Now, if we change the angle $\theta$ to $\pi-\theta$, the correlation 
$C_e$ changes sign. Therefore, by exploiting the relation 
\begin{equation}
d^j_{\lambda_a\lambda_b} (\pi-\theta) = (-)^{j+\lambda_a} 
d^j_{\lambda_a-\lambda_b} (\theta), \label{symm}
\end{equation}
the difference connected to the T-even asymmetry ${\cal A}_e$ reads as
\begin{eqnarray} 
\Delta {\cal P}_e (\theta) &=& {\cal P} (\theta) - {\cal P} (\pi-\theta) 
\label{dpr0}
\\
\ ~~~ \ ~~~ &=& \sum_{\lambda_a, \lambda_b} \sum_{j,j'}
[{\cal A}^j_{\lambda_a\lambda_b}
{\cal A}^{j'*}_{\lambda_a\lambda_b}- (-)^{j+j'+2\lambda_a}
{\cal A}^j_{\lambda_a-\lambda_b}{\cal A}^{j'*}_{\lambda_a -\lambda_b}] 
d^{j}_{\lambda_a\lambda_b} (\theta) d^{j'}_{\lambda_a\lambda_b} (\theta). \label{dpr1}
\end{eqnarray}
Then the overall T-even asymmetry is
\begin{equation}
{\cal A}_e = \frac{\Delta H}{H}, \label{asys}
\end{equation}
with 
\begin{equation}
H =  \int_0^{\pi}d\theta sin\theta {\cal P} (\theta), 
\ ~~~~~~ \
\Delta H = \int_0^{\pi/2}d\theta sin\theta \Delta {\cal P}_e(\theta).
\label{intg}
\end{equation}
By inserting the expressions (\ref{dpr}) and (\ref{dpr1}) into eqs. (\ref{intg}), we get
\begin{eqnarray}
H &=&  \sum_{\lambda_a, \lambda_b}\sum_{j,j'} {\cal A}^j_{\lambda_a\lambda_b} {\cal A}^{j'*}_{\lambda_a\lambda_b} I^{j,j'}_{\lambda_a\lambda_b}, 
\\ 
\Delta H &=& \sum_{\lambda_a,\lambda_b}\sum_{j,j'} 
[{\cal A}^j_{\lambda_a\lambda_b}{\cal A}^{{j'}*}_{\lambda_a\lambda_b}
- (-)^{j+j'+2\lambda_a}{\cal A}^j_{\lambda_a
-\lambda_b}{\cal A}^{j'*}_{\lambda_a -\lambda_b}]
J^{j,j'}_{\lambda_a, \lambda_b}, \label{hdh}
\end{eqnarray}
where
\begin{equation}
I^{j,j'}_{\lambda_a, \lambda_b} =  \int_0^{\pi}
d\theta sin\theta d^{j}_{\lambda_a\lambda_b} (\theta) d^{j'}_{\lambda_a\lambda_b} (\theta), \ ~~~~~~ \
J^{j,j'}_{\lambda_a, \lambda_b} =  \int_0^{\frac{\pi}{2}}
d\theta sin\theta d^{j}_{\lambda_a\lambda_b} (\theta) d^{j'}_{\lambda_a\lambda_b} (\theta). \label{intJ} 
\end{equation}

Some remarks are in order. 

- The T-even asymmetry may be applied to any three-body decay of 
the type 
\begin{equation}
r_0 \to r ~~~~ a ~~~~ b,
\end{equation} 
where $r_0$ and $r$ denote (pseudo-)scalar particles. 

- This asymmetry  is {\it a priori} non-trivial also for a
strong or electromagnetic decay of the type just mentioned.

- It vanishes if the two-particle system ${\cal S}$ has a definite
angular momentum. This can be seen directly by substituting eq. (\ref{symm}) into eq. (\ref{dpr0}). Alternatively, recalling once 
more that ${\cal S}$ is in a helicity eigenstate, we observe that 
no interference occurs in this case, therefore\cite{ajd1} 
${\cal A}_e$ = 0. 

- Our choice of grouping the three final particles according to eq. (\ref{gr}) is arbitrary: two more combinations are possible and all 
of them have to be investigated, in order to single out all possible correlations between particles.

The T-even asymmetry just illustrated may be applied to the search for
new physics. To this end, we suggest to consider the difference 
\begin{equation}
\Delta {\cal A}_e = {\cal A}_e^+ - {\cal A}_e^-, \label{obsv}
\end{equation}
where ${\cal A}_e^{\pm}$ is the asymmetry relative to, {\it e. g.}, a $B^{\pm}$ decay of the type (\ref{dcys}). $\Delta {\cal A}_e$, which
is sensitive to the CP asymmetry of the decay considered, can be seen 
as an effect of time reversal violation, provided the CPT symmetry holds.
In particular, since this asymmetry is T-even - in the sense explained
in refs. \cite{aj,ajd1} - it may be caused by two possible kinds of interference terms. The former consists of the interference between a real time reversal violating amplitude and a fake T-odd one, caused for instance by a spin-orbit interaction\cite{br1,br2}; the latter comes from the interference between a true T-even amplitude and a spin-orbit, time reversal violating one. Obviously, $\Delta {\cal A}_e$ is expected to be zero for strong or electromagnetic decays. For weak decays, the standard model generally predicts small values of this observable; therefore a large value of $\Delta {\cal A}_e$ may be a hint at physics beyond the standard model.
 
\vskip 0.25in
\centerline{\bf Acknowledgments}
The author is thankful to his friends F. Fontanelli and C. Patrignani
for useful and stimulating discussions.

\vskip 1.00cm



\end{document}